\documentclass[accepted]{bmaw2022}  

\usepackage[american]{babel}

\usepackage{natbib} 
\usepackage{mathtools} 
\usepackage{booktabs} 
\usepackage{tikz} 

\usepackage{graphicx}

\title{User-centric Music Recommendations}

\author[1]{\href{mailto:<jaime.ramirez@alu.uclm.es>}{Jaime Ramirez-Castillo}{}}
\author[1]{{\href{mailto:<julia.flores@uclm.es}{M. Julia Flores}{}}}

\author[2]{{\href{Ann.Nicholson@monash.edu}{Ann E. Nicholson}{}}}

\affil[1]{%
    Departamento de Sistemas Inform\'aticos\\
    Universidad de Castilla - La Mancha\\
    Spain
}
\affil[2]{Faculty of Information Technology. Monash University. Clayton. Australia.}

\begin{document}
\maketitle

\begin{abstract}
This work presents a user-centric recommendation framework, designed as a pipeline with four distinct, connected, and customizable phases.
These phases are intended to improve explainability and boost user engagement.

We have collected the historical Last.fm track playback records of a single user over approximately 15 years.
The collected dataset includes more than 90,000 playbacks and approximately 14,000 unique tracks.

From track playback records, we have created a dataset of user temporal contexts (each row is a specific moment when the user listened to certain music descriptors).
As music descriptors, we have used community-contributed Last.fm tags and Spotify audio features. They represent the music that, throughout years, the user has been listening to.

Next, given the most relevant Last.fm tags of a moment (e.g. the hour of the day), we predict the Spotify audio features that best fit the user preferences in that particular moment.
Finally, we use the predicted audio features to find tracks similar to these features. The final aim is to recommend (and discover) tracks that the user may feel like listening to at a particular moment.

For our initial study case, we have chosen to predict only a single audio feature target: \emph{danceability}.
The framework, however, allows to include more target variables.

The ability to learn the musical habits from a single user can be quite powerful, and this framework could be extended to other users.
\end{abstract}

\section{Introduction}\label{sec:intro}
Music recommendation systems have been a popular topic across streaming platforms and the Music Information Retrieval (MIR) research field. These systems have been a fruitful area of research and a common feature implemented, with different degrees of complexity, in many music streaming platforms \citep{ramirez2020machine}.

Music recommendations, however, are not always easy to interpret from the perspective of the listener. Users often get recommendations without clear, meaningful explanations or justifications, such as why the listener might like a specific track more than others.

One solution to this problem is to explore recommender systems that focus on the users' individual habits and their listening contexts. User-centric recommendation approaches are a response to the traditional, generic machine learning approaches focused on reproducing ground truth and maximizing accuracy.
Unlike generic techniques, user-centric mechanisms aim to develop systems built on top of the user context \cite{schedl2013neglected}.

In this paper, we set the foundations of a user-centric music recommendation framework. To us, the user-centric aspect should not only consider specific circumstances about the user context, but also give users enough feedback and control to enrich the recommendation process and the listening experience.

The remainder of the article is organized as follows. In section 2, we introduce the foundational ideas behind our user-centric recommender system. Subsequently, section 3 covers the methodology used to gather the data, prepare the data, and build the recommendation pipeline. Section 4 demonstrates how the recommendation pipeline works. Finally, in section 5, we discuss our conclusions and future work possibilities.

\section{A User-centric Recommendation Framework}

In recent years, music recommendation algorithms have been heavily influenced, improved, and driven by the progress made in Artificial Intelligence (AI) and Machine Learning (ML).
One aspect of many modern AI/ML algorithms, however, is their black-box nature.
It is difficult for users to ask \emph{Why} and \emph{How} a model is producing a particular output.

Although researchers are making progress with explainability techniques \citep{zhang2020explainable}, we believe that music recommendations can be improved beyond explainability.

We envision a process similar to what music aficionados might experience when they enter a record store, ask for professional advice, and listen to a number of records.
It might be the case that the music store owner even knows the preferences of the customer, and recommends records accordingly.

We are also interested on exploring how an intelligent system might be able to reproduce scenarios where a single user picks tracks for listening, based on the user habits.
The focus is not to generalize to other users, but to pick a track, or a record, that a single user would like to listen to at a given moment.
This approach might be a way to, based on a single user history of music listening habits, make generalizations within the scope of the user, and with the perspective of the user.

With this idea in mind, we have designed a music recommendation framework as a pipeline that processes the listener's preferences, context, and music metadata, to produce recommendations.

To prevent the black-box effect, we have split the recommendation process into a series of explicable steps, which we call \emph{phases}. The idea, with each one of these phases, is to reinforce the user engagement, by gathering results, inspecting explanations, and providing preferences in each phase.

\subsection{A Single-user Dataset}

Similar to other intelligent systems, recommender systems must be trained, by using user preference data, to produce adequate recommendations.
For our recommendation framework, we have leveraged the knowledge discovery potential of large historical listening logs, gathered from Last.fm.

To characterize the preferences and context of the user, we have chosen to start with a simple scenario, where just data from a single user is available.
By training our system with data from a single user, we also want to begin a discussion, given the following question: \emph{Is it possible to train recommender systems, and in particular, user-centric systems, by using a single-user dataset?}

To the best of our knowledge, research on user-centric recommender systems has concentrated its efforts on explainable AI \cite{wang2019designing}, and also user-centered evaluation of these systems \cite{knijnenburg2012explaining}. We also argue that recommender systems that exploit the preferences of a single user, or a reduced number of users, might as well be considered as user-centric models.

\subsection{User Listening Context Data}

We have generated our single-user dataset by using the Last.fm listening history of a single user.
Other than the playback timestamp, the Last.fm API does not provide any other additional details about the user's context \footnote{Further details about the Last.fm API can be found at \url{https://www.last.fm/api}}.
Therefore, the only context we have available is the moment when the user played each track.
These moments, available in the Last.fm API as timestamp objects, provide us with the following temporal context:

\begin{itemize}
\item Track playback time, including hour, and minute.
\item Track playback date, including day, month, and year.
\end{itemize}

\subsection{Track Characterization Data}

To characterize the music preferences of the user, we have gathered the following data:

\begin{itemize}
\item \emph{Community-contributed tags from the Last.fm API}. These tags are text labels that Last.fm users assign to artists, albums, or tracks.
Users apply these tags to categorize music from their own perspective, which means that tags do not fit into any structured ontology or data model. Tags can refer to aspects such as genre, emotion, or user context.
\item \emph{Track audio features from the Spotify API}. These are attributes computed from the audio themselves. They are a way to describe music by using numerical values. For example, a \emph{danceability} attribute of 0.95 means that a particular song is highly suitable for dancing.
\end{itemize}

\begin{figure}
  \centering
  \includegraphics[width=0.95\linewidth]{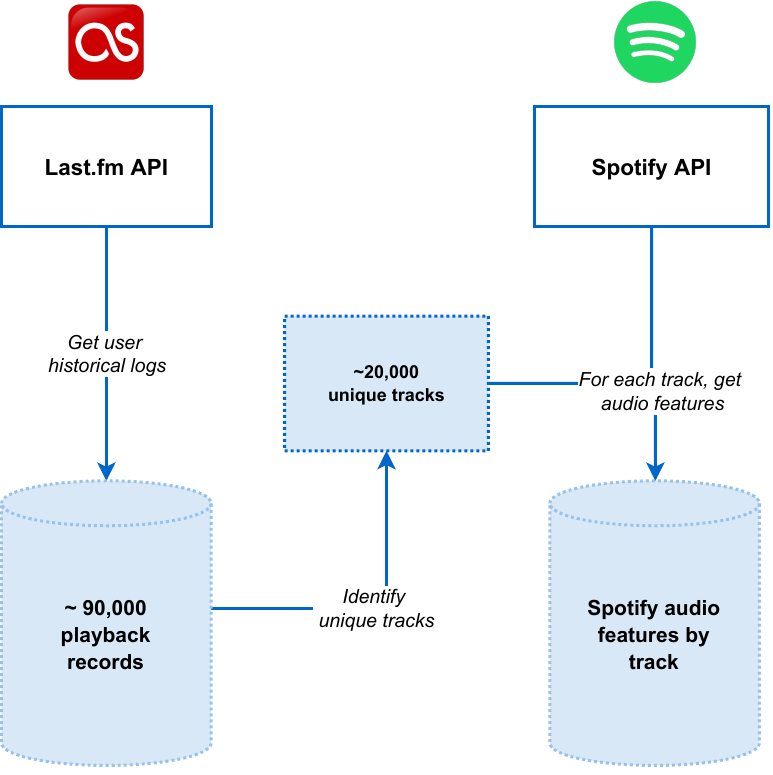}
  \caption{\textbf{Data gathering}. We have collected the historical logs of a single user from Last.fm, for the 2007-2021 period. The data includes more than 90,000 track playback records. For each track the user has ever listened to, we have gathered Spotify audio features. }\label{fig:data_gathering}
\end{figure}

To limit the scope of our data gathering efforts, we have only collected the Last.fm tags and Spotify audio features of the tracks that the user has listened to, as shown in figure~ \ref{fig:data_gathering}.

The reader might also want to note that the Spotify API provides low-level track analysis data, which we have not leveraged in this study\footnote{
    Further details about the Spotify Web API can be found at \url{https://developer.spotify.com/documentation/web-api/}.
}.

\subsection{The Phases of the Recommendation Framework}

The following items are the phases the recommendation framework is built upon.
These phases only describe the inference process, after the data has been preprocessed and the models have been trained.

\begin{itemize}
\item \emph{Phase one}: When the user starts the listening session, this phase suggests the most suitable Last.fm tags to listen to.

This phase aggregates the strength of Last.fm tags in a specific moment, given the user listening records. This value is computed for the most relevant (top-k) Last.fm tags found in the user listening history. For example, if, throughout the years, the user has been consistently listening to relaxing music before going to sleep around 11:00 PM, then the tag \emph{Relaxing} might show a high strength value at this hour.

\item \emph{Phase two}: Generate the \emph{most-attractive fake track}, represented as a list of Spotify audio features.

In the second phase, a regression model receives the list of Last.fm tag strength values as the input, and predicts the values of the Spotify audio features.
The model output is a list of Spotify audio features, structurally equal to what we would get if we queried the Spotify API to get the features of any real track.
Therefore, we interpret this output as the hypothetic, most-likable, fake track, which is what the model believes that the user wants to listen to, given a set of Last.fm tags.

\item \emph{Phase three}: Find real tracks closest to the prediction.

A ranking system looks for tracks that are similar to the fake track, by computing the distance of the Spotify audio features of each of of these tracks with the fake track.
We have limited the scope of this query by selecting only tracks that are in the user track collection.

\item \emph{Phase four} (not developed): Our plan is to post-process the recommendations by applying diverse techniques and criteria, allowing the user to participate in this process, e.g. by selecting the degree of exploration versus exploitation
\cite{dingjan2020exploring}.

\end{itemize}

\section{Methodology}

The methodology here presented describes the details of the following processes:

\begin{itemize}
\item Data gathering.
\item Data preparation and aggregation.
\item Phase-two model training.
\item Inference: four-phase recommendation pipeline.
\end{itemize}

\subsection{Data Gathering}

We have built a dataset tailored to the input requirements of the Phase-two regression model.
These requirements can be defined as follows:

\begin{itemize}
\item Input parameters ($X$): Each sample consists of a list of Last.fm tag strength values, which describe how relevant each Last.fm tag is for each particular sample.
Each sample maps to a particular moment in the users' listening history (e.g. 11 AM, August 15th, 2012).

\item Output values ($y$): The most suitable values of the Spotify audio features for a specific moment.
\end{itemize}

To construct the dataset, we have downloaded the data from the Last.fm and Spotify APIs.
The user we picked for our research has been sending telemetry data to Last.fm since 2007.
This user has reported more that 90,000 track playbacks over 15 years\footnote{
    The Last.fm account used in this work belongs the corresponding author of this article.
    The listening history of this user is available at \url{https://www.last.fm/user/jimmydj2000/}.
}.

\subsubsection{Last.fm Tags}

Last.fm uses the term \emph{scrobble} to refer to a single track playback, in a particular moment.
We have queried the Last.fm API to download the user's scrobble logs, reported from 2007 to 2022.
For each scrobble, we have gathered the following information:

\begin{itemize}
\item Track playback timestamp.
\item Track MusicBrainz Identifier (MBID), if exists.
\item Track name
\item Artist name
\item Track tags. If the track does not have any tags assigned, then artist tags have been used.

\end{itemize}

For each tag assigned to a track, or an artist, Last.fm includes a \emph{count} property to indicate the popularity, or strength, of the given tag for the track.
This value is normalized from 0 to 100, so the most popular tag (for a track) can have a count value of 100.

Users normally listens to their favorite tracks many times, so the amount of individual tracks listened is much smaller than the number of track plays.
In this case, the amount of individual tracks listened is about 20,000.

\subsubsection{Spotify Audio Features}

After gathering Last.fm data and identifying the unique tracks that represent the user music collection, we have collected Spotify audio features.
For each of these individual tracks, we have downloaded the Spotify audio features specific to the given track.

The Spotify audio features are numerical values that represent high-level audio information computed from a specific track.
These values characterize a track, musically speaking, by measuring relevant musical aspects.

The features provided by the Spotify API are: \emph{acousticness}, \emph{danceability}, \emph{duration\_ms}, \emph{energy}, \emph{instrumentalness}, \emph{key}, \emph{liveness}, \emph{loudness}, \emph{mode}, \emph{speechiness}, \emph{tempo}, and \emph{valence}.
Table \ref{tab:spotify-features} describes these features.
The reader can find further details about each feature in the Spotify API documentation
\footnote{
    See \url{https://developer.spotify.com/documentation/web-api/reference/get-audio-features}.
}

A small portion of the tracks do not have features available in Spotify, so they have been filtered out from our experiments.

\begin{table}
    \centering
    \caption{Spotify audio features. These features provide high-level musical information about a track.} \label{tab:spotify-features}
    \begin{tabular}{p{0.3\linewidth}p{0.6\linewidth}}
        \toprule
        \bfseries \textbf{Feature name} & \textbf{Description} \\ 
        \midrule
        \textbf{acousticness} & The track is acoustic. From 0 to 1 \\
        \textbf{danceability} & The track encourages (or is adequate for) dancing. From 0 to 1 \\
        \textbf{duration\_ms}  &  Duration in milliseconds \\
        \textbf{energy}  &  The track is perceived as energetic. From 0 to 1\\
        \textbf{instrumentalness}  &  The track is instrumental. From 0 to 1 \\ 
        \textbf{key}  &  Key categories encoded as integers. From C (0) to 11 \\ 
        \textbf{liveness}  &  The audience is audible. From 0 to 1\\
        \textbf{loudness}  &  In decibels. From -60 to 0 \\ 
        \textbf{mode}  & Major (1) or minor (0) \\ 
        \textbf{speechiness}  & Does the track contain speeches? From 0 to 1 \\
        \textbf{tempo}  & In beats per minute (BPM) \\ 
        \textbf{valence} & How happy is the track (BPM). From 0 to 1 \\ 
        \bottomrule
    \end{tabular}
\end{table}

\subsection{Data Preparation}

After having collected the data from Last.fm and Spotify, the unprocessed dataset included, approximately, 90,000 playback records of 20,000 individual tracks, which correspond to the listening activity of a single user throughout years.
Each track is associated with a list of Last.fm tags (and their strength), and a list of Spotify audio features.

\subsubsection{Last.fm Tags Reduction}

Counting the total amount of Last.fm tags in the user collection resulted, initially, in more that five million tags.
We quickly confirmed that building a tabular data set, in which every row contains millions of columns (Last.fm tags) was doable, but presented scalability problems.

Therefore, we decided to reduce the number of tags by picking a subset of the most relevant tags.
The reduction algorithm is simple: calculate a weighted sum of all the tags appearances and pick the top 1000.
The sum is weighted because we use the \emph{count} attribute.
This attribute is present in every Last.fm track-tag association and provides a measure of the strength of a particular tag in a specific track.

Note that this reduction is an initial approach, which, similar to other phases, can be extended or improved in the future.
For this particular case, dimensionality reduction algorithms, such as PCA, are good candidates for forthcoming iterations of this work.

\subsubsection{Reducing the Number of Samples}

Our initial intention was to generate a dataset of musical moments, or intervals.
Moments (e.g 2007-07-31, from 6:00 PM to 7:00 PM) when the user listened to at least one track, since 2007.
So, rather than using each track playback as a data record for training, we have grouped the data by \emph{Year-Month-Day-Hour} intervals.

Next, for each interval, we have aggregated the strength values of the top-1000 Last.fm tags and Spotify audio feature values.
Once again, we decided to choose the simplest approach and calculate the mean to aggregate these values.

The following algorithm describes this grouping and aggregation process in further detail:

\begin{enumerate}
\item Group track playbacks by \emph{Year-Month-Day-Hour} intervals.
\item For each interval $i$:
\begin{enumerate}
\item For each tag $t$, of the top-1000 Last.fm tags, found in the interval, calculate $tagIntervalStrength_{i_t}$ as the sum of all the strength values of all the appearances of tag $t$ in interval $i$.
E.g., if, from 5 PM to 6 PM, the tag \emph{rock} happens twice, with strength values of 100 and 80, then the total strength of \emph{rock} for the interval is 180.
\item Calculate $totalStrength$ as the sum of all tag strength values $tagIntervalStrength_{i_t}$ found in interval $i$.
\item For each $t$, divide $tagIntervalStrength_{i_t}$ by $totalStrength$.
\end{enumerate}
\end{enumerate}

This algorithm generates the relative frequencies (weighted by strength) of the top-1000 Last.fm tags happening in \emph{Year-Month-Day-Hour} intervals. We have normalized the result so that the values of these 1000 tags sum up to 100 for each interval. These are not, strictly speaking, probabilities, because they range from 1 to 100, and because how they have been constructed. We could, however, interpret them as probabilities, just dividing them by 100 and making them belong to the range [0.0,1.0]. 

In the same direction, if we have, as evidence, that the current hour ($H$) is 17:00, these values could be interpreted as $P(tag_i,...,tag_k|H=17{:}00)$. We could argue that we have a conditional probability distribution where we consider the probability (in fact, the relative frequency) of the k top 1000 tags, given a particular moment.  

\subsubsection{Spotify Features Reduction}

Just like we have done with Last.fm tags, we have also generated average, or aggregated, values of Spotify features for each \emph{Year-Month-Day-Hour} interval.
To aggregate the values of Spotify features, we have grouped track playbacks by \emph{Year-Month-Day-Hour} interval, calculated the mean value, per interval, of each feature.

\subsubsection{The Resulting \emph{Musical-moments} Dataset}

The resulting \emph{musical moments} dataset is comprised of 14,203 samples.
Each sample contains 1,000 Last.fm tag strength values, which sum up to 100, and 12 Spotify audio features.

By reducing the data we have initially collected, we have converted more than 90,000 track playback samples, into a dataset of 14,000 \emph{musical moments}.
Additionally, with this reduction, we have been able to minimize the hardware requirements and the training time required for the initial experiments of our framework.
Finally, by training our system with \emph{Year-Month-Day-Hour} moments, we have created a system that can deliver recommendations by just taking the hour of the day as the input.

Also, as explained earlier, we have designed the framework with customization in mind.
Using \emph{Year-Month-Day-Hour} intervals, or partitions, is our initial approach, but a different aggregation strategy might be chosen for future experiments.

\begin{figure}
  \centering
  \includegraphics[width=\linewidth]{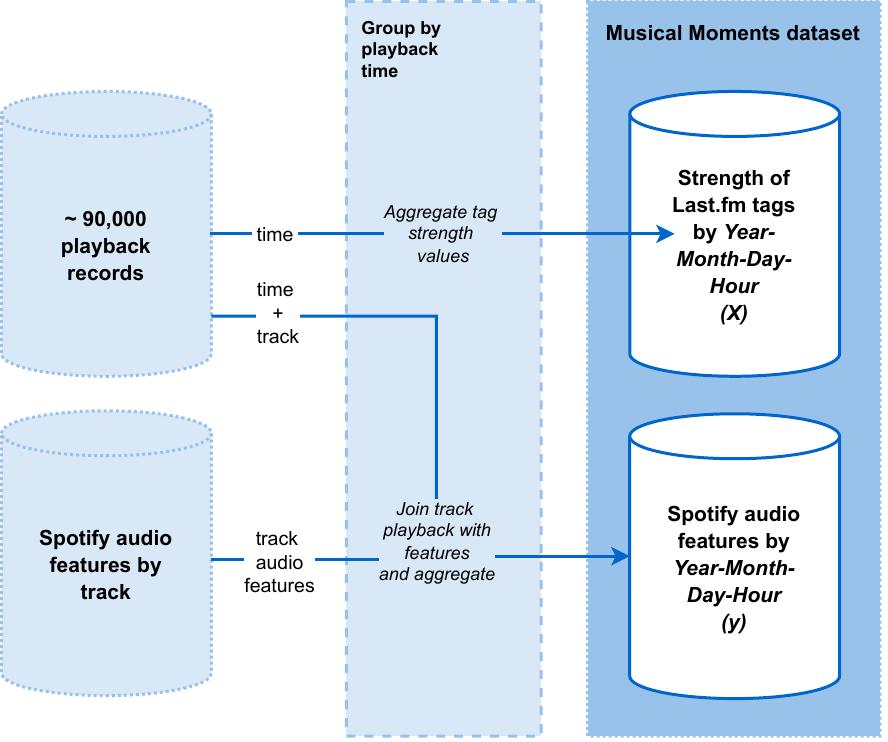}
  \caption{Data preparation.}\label{fig:data_prep}
\end{figure}

\subsection{Phase-two Model Training}

After data preparation, the resulting \emph{musical moments} dataset contains 14,203 samples.
Each sample includes 1,000 columns that correspond to Last.fm tag strength values, and 12 columns that correspond to Spotify audio features. 

The Spotify features are separated from the Last.fm tag values, so a \emph{timestamp} field is used to index and join both sets. 
The Last.fm data file looks like the following example:

\begin{verbatim}
timestamp     electronic seen live  ...
2022-03-28T17 1.01822    7.28831    ...
...
\end{verbatim}

Likewise, the Spotify features, indexed the same temporal intervals, look as follows:

\begin{verbatim}
timestamp     acoustic. danceab. ...
2022-03-28T17 0.05211   0.6085   ...
...
\end{verbatim}

By using these data, we can now train the model, which is intended to be used in Phase two, to predict Spotify features ($y$), given a list of 1000 Last.fm tag strength values ($X$).

When training, we have removed the timestamp column index, so that the model does not have any sense of time.
We also split the dataset into training and validation sets.

For the sake of simplicity, our initial experiment in Phase two trains the model to predict only one Spotify feature: \emph{danceability}.

\begin{figure}
  \centering
  \includegraphics[width=\linewidth]{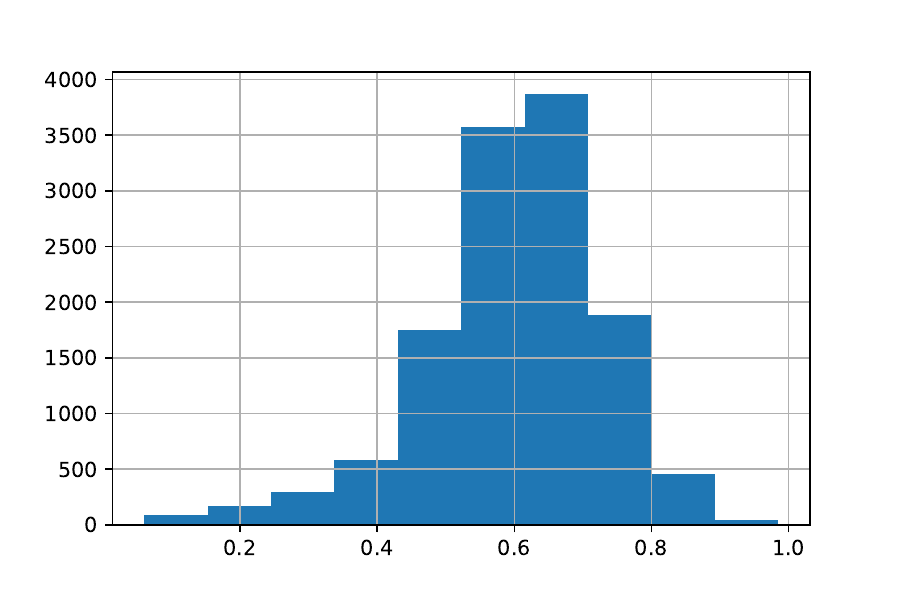}
  \caption{Distribution of the \emph{danceability} Spotify audio feature across 20,000 individual tracks, which correspond to the single-user dataset used in this paper.}
  \label{fig:danceability}
\end{figure}

By looking into all the \emph{danceability} values we have gathered for 20,000 tracks, we can see that the mean value of the \emph{danceability} feature is 0.599, and the standard deviation is 0.13, as figure \ref{fig:danceability} depicts.


\subsubsection{Training Experiments and Results}

As already mentioned, the framework has been designed to be open for extension and customization. This means that the strategies applied to each of the framework phases should be replaceable and customizable.
Because our intention is to validate the framework, rather than an individual model, our training experiments have been simple.
We have conducted a round of simple tests by just training two models. Three, if we count the baseline model. 

The \emph{musical moments} dataset has been split into a training set and a test set, generating a training set with 8482 samples, and a test set with 4179 samples.
After training each model on the training set, we have evaluated the model on the test set, by using the RMSE metric.

Initially, we have defined our baseline model as a simple, random predictor.
This model generates predictions by following a normal distribution defined by the mean and the standard deviation of the \emph{danceability}.

Next, we have trained an XGBoost regression model and a Bayesian ridge model.
Table \ref{tab:training-results} shows the results of these experiments.
The experiment with XGBoost has produced the lowest RMSE, so we have decided to use this model for the recommendation pipeline.

\begin{table}
    \centering
    \caption{\textbf{Phase-two model training experiments}. The RMSE of each model, as well as the training time, in seconds. Each model has been trained on a training set of 8482 moments. The RMSE value is the result of using the 4179 moments of the test set to evaluate the model.} \label{tab:training-results}
    \begin{tabular}{lll}
        \toprule
        \bfseries \textbf{Model} & \textbf{RMSE} & \textbf{Seconds} \\ 
        \midrule
        \textbf{Baseline} & 0.19 & ~0 \\ 
        \textbf{XGBoost} & 0.09 & 19 \\
        \textbf{Bayesian Ridge} & 0.10 & 2 \\
        \bottomrule
    \end{tabular}
\end{table}

\subsection{Running the Recommendation Pipeline}

After preparing the data and training the Phase-two model that predicts Spotify features from Last.fm tag strengths, we have built the phased pipeline for our music recommendation framework.

For each phase, we detail the input and output values, as well as how the phase can be explained to the user.

\begin{figure}
  \centering
  \includegraphics[width=1\linewidth]{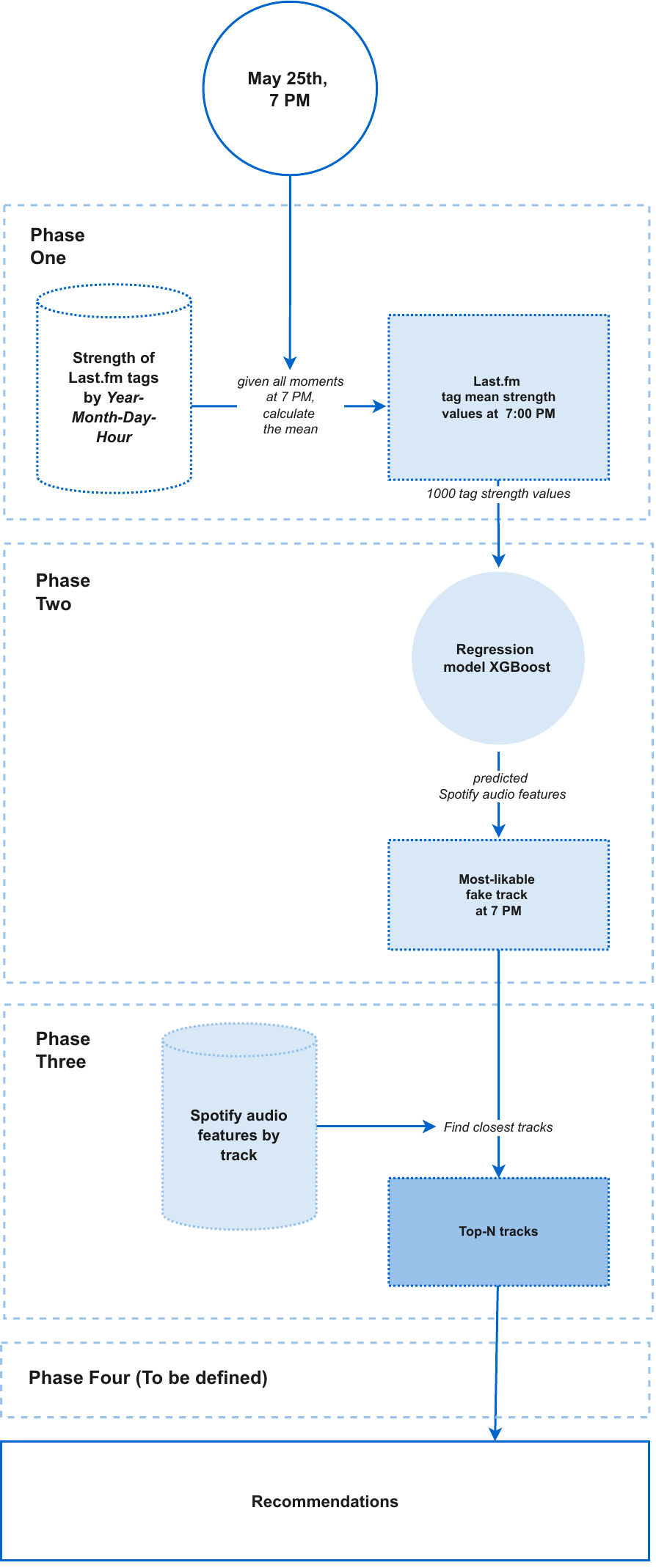}
  \caption{\textbf{Recommendation pipeline}. The recommendation pipeline begins with a particular moment in time (the hour of the day). Phase one aggregates the available Last.fm tag strength values for this particular hour. Phase two predicts the most-suitable Spotify audio features for this hour, and finally Phase three selects the tracks that are closest to this prediction. The diagram also depicts Phase Four, to provide an idea of where this additional phase fits in the framework.}
  \label{fig:pipeline}
\end{figure}

\subsubsection{Phase One: Estimate Current-moment Last.fm Tag Strength Values}

\begin{itemize}
\item \emph{Input}: the current hour.
\item \emph{Output}: Estimated strength of the top-1000 Last.fm tags for the current hour.
\item \emph{Explainability}: The outputs of this phase can be explained  as $ P(Tag_1...Tag_n | Hour) $
\end{itemize}

Phase one determines the current local time of the listener and extracts the hour.
Given the hour, our system generates the probabilities of 1000-Last.fm-tag set.

The simplest solution for us has been to group the 1000-Last.fm-tag set by hour, calculate the average for each hour, and select the current hour.

Phase one aims at meeting the following goals:
\begin{enumerate}
\item Generate the top-1000 Last.fm tag strength values to feed into phase two.
\item Provide the listener with insight about what are the system beliefs about the user preferences in a certain moment.
\end{enumerate}

\subsubsection{Phase Two: Predict Spotify Features}

\begin{itemize}
\item \emph{Input}: Top-1000 Last.fm tag strength values.
\item \emph{Output}: Predicted Spotify audio features.
\item \emph{Explainability}: We have not implemented any explainability strategy for this phase. However, the predictions of any model applied to this phase could be explained, either because the model is a \emph{white-box}, interpretable model (such as the Bayesian model), or because explainability practices are put into practice.
In this case, the model we use is XGBoost, which is difficult to interpret. For this model, and other \emph{black-box} models, the predictions can be explained with techniques such as LIME  \cite{ribeiro2016model}.
\end{itemize}

Phase two uses the Last.fm strength values compiled in the previous phase and feeds these values into the model to predict Spotify feature values.

\subsubsection{Phase Three: Ranking}

\begin{itemize}
\item \emph{Input}: Spotify audio features. For the experiments of this paper, we only include the \emph{danceability} feature.
\item \emph{Output}: A top-k set of tracks, ordered by descending distance to \emph{danceability}. 
\item \emph{Explainability}: The ranking only takes the danceability feature into account and selects the tracks with the closest danceability value. 
\end{itemize}

By using the predicted Spotify audio features, we select tracks that are closest to the predicted features. Basically, the system proposes a set of songs, which are close to the desired audio feature in a particular moment.
For the scenario we are explaining in this paper, we have only computed the distance to the \emph{danceability} feature, but we could compute this distance to any other target variables, or even a combination of them.

Tracks ranking, or selection, can be performed on any collection of tracks if the Spotify features for those tracks are known.
In our case, we have queried the dataset of Spotify audio features that we initially gathered for 20,000 unique tracks, which can be interpreted as the music collection of the user.

\subsubsection{Phase 4: Fine-tuning}

This phase has not been developed yet.

We intend to post-process the results, based for example, on the degree of exploration, decided by the listener. The input of this phase is a ranked list of the top-M songs closer to the target feature. The user may want to include later some kind of extra requirement, that could be considered in this phase.

\section{Experimentation}

The following listings demonstrate how running the recommendation pipeline on a very rudimentary terminal-based user interface can provide recommendations, as well as feedback to the user.

\subsubsection{Run recommendation pipeline at 19:00 PM}

\begin{verbatim}
PHASE 1: Compute Last.fm Tags
  · Current Time: 19:23
  · Tag strength at 19:00-20:00
    | electronic | electronica | ...
    | 10.945045  | 3.170302    | ...

PHASE 2: Predict Spotify Features
  · Danceability: 0.58332
  
PHASE 3: Ranking - Closest Tracks
  1. Kelly Lee Owens - Jeanette
     · danceability: 0.583
     · distance: 0.0003215

  2. ...
\end{verbatim}

Phase three returns 20 tracks.
The user has listened to 11 tracks and skipped the other nine.

\subsubsection{Run recommendation pipeline at 9:00 AM}

\begin{verbatim}
PHASE 1: Compute Last.fm Tags
  · Current Time: 9:25
  · Tag strength at 9:00-10:00
    | electronic | electronica | ...
    | 11.202752  | 3.096055    | ...

PHASE 2: Predict Spotify Features
  · Danceability:  0.620887

PHASE 3: Ranking - Closest Tracks
  1. Wiki & Oneplus - Rise to the Surface
     · danceability: 0.621
     · distance: 0.0001126

  2. ...
\end{verbatim}

The user has listened to 12 tracks.
The other eight tracks have been skipped.

\section{Discussion and Future work}

This paper reflects our initial efforts to define a user-centric, explainable, music selection and recommendation framework. We have devised a modular scheme that can be used in distinct scenarios, as certain decisions can be changed within the phases but the design of the recommendation pipeline makes sense equally. 

This work has described the main phases of the framework, and has provided a study case where we have worked with the history log of a single user. We have created a personalized recommended system, which could be applied to any other users who have their track playback history available.

Apart from its clear application to other datasets (with one or more users), the strongest point of this framework is the versatility, not only in the determination of certain parameters, and the models to be learned, but also in the way tags are chosen, or in the target features. In the future, we would like to explore this approach by trying to find the best uses and configurations.

We state that the framework is modular because phases are established in a high-level and abstract way, and we could actually inject diverse strategies or algorithms in each phase. For example, in phase one, our study case has grouped listened tracks by hour. However, different time perspectives could have been used (morning/afternoon/evening, date+hour, weekday vs holidays/weekend, music that the user listen on a Saturday morning, seasons, etc..). Also, the aggregation functions to compute Last.fm tag strength values are simple \emph{mean} functions, but this can grow in complexity if we use more complex estimation functions, or even more advanced techniques, such as clustering. 

With respect to Phase two, we have presented a regression problem for a single predictive variable. However, there is much room to experiment, both in the prediction problem and in the paradigm of machine learning models. We can, as well, explore explainability techniques here.

In this initial study case, we have just used the \emph{danceability} audio feature, but there are, as seen in Table \ref{tab:spotify-features} other interesting values provided by Spotify. We could have used any of them, and we think a combination of the most relevant ones could be a promising strategy, which would transform the problem into a multi-target regression task\cite{xu2019survey}.

When coping with Phase 3, we could apply more advanced ranking algorithms. If, for example, we could predict distinct target variables, how to generate the ranking of closer tracks would imply more sophisticated techniques, where specific multi-dimensional distances should be applied \cite{liu2018metric}.

Finally, Phase 4 is still in an inception stage, and is open to post-processing options, which could be given to the user interactively, or could imply a finer selection of the proposed tracks. In this phase, we plan to experiment with exploration vs exploitation trade-offs.

In summary, this paper presents a new framework oriented to user-centered recommendations in music listening. Also, we have described a particularization of it, where we can provide the user with a set of recommended songs, given the time of the day, and based on a large historical track playback log. Collecting, preprocessing and combining data from Last.fm and Spotify is also a contribution, as this have been done explicitly for this work. The applicability of the current work and its possible extensions is clear and quite straightforward. We plan to improve, extend and make further and experiments in the near future.







\bibliography{userrecomm}

\begin{contributions} 

    J.~Ramirez-Castillo and M. Julia Flores constructed the framework idea and wrote the central parts of the paper.
    J.~Ramirez-Castillo collected the data and combined the distinct sources of information to create the dataset. He made most of the coding, aided in the process by M.~J.~Flores and A.~E.~Nicholson.
    A.~E.~Nicholson contributed to the paper revising the methodology and presenting her ideas to improve the framework. She also contributed to the writing of the paper. 
\end{contributions}

\begin{acknowledgements} 
   This work has been partially funded by JCCM (Junta de Comunidades de Castilla - La Mancha), FEDER funds and the Spanish Government through projects  SBPLY/21/180501/000148 and PID2019-106758GB-C33.

\end{acknowledgements}

\end{document}